\documentclass[AMA,STIX1COL,12pt]{WileyNJD-v2}
\usepackage{moreverb}
\usepackage{mathtools}

 \newcommand{\ind}{\perp\!\!\!\!\perp} 

\newcommand\BibTeX{{\rmfamily B\kern-.05em \textsc{i\kern-.025em b}\kern-.08em
T\kern-.1667em\lower.7ex\hbox{E}\kern-.125emX}}

\articletype{Research Article}%

\received{<day> <Month>, <year>}
\revised{<day> <Month>, <year>}
\accepted{<day> <Month>, <year>}

\begin{document}

\title{A Novel Causal Mediation Analysis Approach for Zero-Inflated Mediators}

\author[1]{Meilin Jiang}
\author[2,3]{Seonjoo Lee}
\author[4,5]{A. James O'Malley}
\author[6]{Yaakov Stern}
\author[1]{Zhigang Li*}

\authormark{Jiang \textsc{et al.}}

\address[1]{\orgdiv{Department of Biostatistics}, \orgname{University of Florida}, \orgaddress{\state{Gainesville, FL}, \country{USA}}}

\address[2]{\orgdiv{Mental Health Data Science}, \orgname{New York State Psychiatric Institute}, \orgaddress{\state{New York, NY}, \country{USA}}}

\address[3]{\orgdiv{Departments of Biostatistics and Psychiatry}, \orgname{Columbia University}, \orgaddress{\state{New York, NY}, \country{USA}}}
\address[4]{\orgdiv{Department of Biomedical Data Science}, \orgname{Geisel School of Medicine at Dartmouth}, \orgaddress{\state{Hanover, NH}, \country{USA}}}
\address[5]{\orgdiv{The Dartmouth Institute}, \orgname{Geisel School of Medicine at Dartmouth}, \orgaddress{\state{Hanover, NH}, \country{USA}}}
\address[6]{\orgdiv{Department of Neurology}, \orgname{Columbia University}, \orgaddress{\state{New York, NY}, \country{USA}}}

\corres{*Zhigang Li, Department of Biostatistics, University of Florida,  Gainesville, FL, USA. \email{zhigang.li@ufl.edu}}

\presentaddress{Department of Biostatistics, University of Florida,  Gainesville, FL, USA}

\abstract[Abstract]{
Mediation analyses play important roles in making causal inference in biomedical research to examine causal pathways that may be mediated by one or more intermediate variables (i.e., mediators). Although mediation frameworks have been well established such as counterfactual-outcomes (i.e., potential-outcomes) models and traditional linear mediation models, little effort has been devoted to dealing with mediators with zero-inflated structures due to challenges associated with excessive zeros. We develop a novel mediation modeling approach to address zero-inflated mediators containing true zeros and false zeros. The new approach can decompose the total mediation effect into two components induced by zero-inflated structures: the first component is attributable to the change in the mediator on its numerical scale which is a sum of two causal pathways and the second component is attributable only to its binary change from zero to a non-zero status. An extensive simulation study is conducted to assess the performance and it shows that the proposed approach outperforms existing standard causal mediation analysis approaches. We also showcase the application of the proposed approach to a real study in comparison with a standard causal mediation analysis approach.
}



\maketitle


\section{Introduction}

Mediation analysis is a statistical technique for investigating causal pathways that are mediated by one or more intermediate variables.\cite{mackinnon2007mediation,VanderWeele2009,Imai2010} Typically, there are at least three variables involved in a mediation analysis: an independent variable which could be an exposure variable in observational studies or treatment in clinical trials, a mediator and an outcome. The total effect of the independent variable on the outcome can be decomposed into indirect effect (i.e., mediation effect) and direct effect, where the indirect effect is the effect attributed to the change in the mediator resulting from the change in the independent variable and direct effect is the effect of the independent variable while holding the mediator unchanged. Both the mediation effect and the direct effect can be estimated and tested in a mediation analysis although the mediation effect is usually the focus. Traditional mediation analysis approaches \cite{mackinnon2007mediation, Judd1981,Sobel1982,Baron1986} are based on linear regression models where the mediation effect is usually estimated by the product of coefficients of the exposure (or treatment) on the mediator and the mediator on the outcome given the exposure. Those approaches are commonly seen in psychology and social sciences, and have been widely applied in other areas as well. Due to the linearity assumption, traditional approaches cannot deal with nonlinear mediation effects or interactions between the independent variable and the mediator. In recent decades, a more flexible mediation analysis framework has been developed based on counterfactual outcomes (i.e., potential outcomes) \cite{Robins1992, Pearl2001} and this approach can allow exposure-mediator interactions and nonlinearities. 
VanderWeele \cite{VanderWeele2009} developed the marginal structural models (MSM) under the counterfactual framework for mediation analysis where marginal models for the outcome and the mediator are specified and natural indirect effect (i.e., mediation effect) and natural direct effect are defined using counterfactual variables. Formulas for the mediation effect and direct effect can be derived based on the marginal models to make inference. Imai et al \cite{Imai2010} developed another approach (referred to as IKT herein) for causal mediation analysis under the counterfactual framework where simulation is used to generate counterfactual variables to estimate the mediation effect. In order to make causal inference, both MSM and IKT approaches together with the traditional approaches require the assumptions of no unmeasured confounding in the models for the outcome and the mediator.\cite{Imai2010,Vanderweele2015} Our proposed approach will be based on the the counterfactual-outcome framework and requires these assumptions as well. In the literature, there have also been approaches \cite{Zheng2014,Guo2018} developed for causal mediation analysis where some no-unmeasured confounding assumptions can be relaxed and replaced by an alternative set of assumptions. As mediation analysis becomes increasingly popular, various extensions have been developed to analyze complex situations such as time-to-event data, time-varying variables, and multiple and high-dimensional mediators.\cite{VanderWeele2016}

Nevertheless, challenges arise in mediation analyses when there are an excessive number of zero-valued data points for mediators because distributional assumptions may be violated due to excessive zeros that could include true zeros and false zeros. In our motivating example, for instance,  the mediator is the brain lesion count known as white matter hyperintensities (WMH). The WMH is a common measurement on a brain damage from magnetic resonance imaging (MRI) scan in the central nervous system caused by diseases like multiple sclerosis. In neuroscience, WMH are the areas that show the increased brightness in MRI images, such as fluid attenuated inversion recovery due to subtle changes in water contents.\cite{wardlaw2015white} WMH is commonly observed in the elderly and is a risk factor of cardiovascular and neurodegenerative diseases.\cite{habes2016white,merino2019white} Recently, studies showed that WMH is a core feature of Alzheimer's disease \cite{lee2016white} beyond a reflection of cerebral amyloid angiopathy.\cite{lee2018white} Thus, there is an increasing demand in studying the role of WMH in cognitive aging \cite{moura2019relationship} and neurodegenerative disease. These relationships make WMH a potential mediator to transmit effects of aging on cognitive abilities or general intelligence. However, WMH often includes a larger number of zeros since it can be rare for a whole sample of participants to have such injury or scarring on a specific lobe of their brains at the time of the scan.\cite{Francois2012} The lesion counts could also be too less or too scattered to be detected depending on the accuracy and precision of the medical equipment. That poses challenges when performing mediation analysis due to the violation of statistical assumptions. It is crucial to recognize and account for the zero-inflated structures of these mediator variables, but there is an unmet need for mediation analysis methods to deal with mediators with zero-inflated distributions. 

To address the unmet need, we propose a new approach to estimate and test mediation effects of zero-inflated mediators with excessive zeros under the counterfactual-outcomes framework. Motivated by our real data example, zero-inflated log-normal (ZILoN) and zero-inflated negative binomial (ZINB) distributions will be the assumed distribution of mediators in the population in the model for the mediator given the exposure. We also considered the zero-inflated Poisson (ZIP) distribution for mediators, a special case of ZINB. In addition, ZINB is capable of handling over-dispersion in the count data. As a result of the zero-inflated nature, the mediation effect can be decomposed into two components. The first component is the mediation effect through the numeric change of the mediator on its continuous scale and this mediation effect is a sum of two causal pathways, and the second component is the mediation effect through only the binary change of the mediator from zero to a non-zero status.\cite{wuMediation} The excessive zeros could be a mixture of true zeros and false zeros because a positive value may be observed as a zero due to technical errors or a limit of detection. The probability of observing false zeros is modeled by a probabilistic mechanism that depends on the underlying true value of the mediator variable.  

This paper is organized as follows. The real dataset that motivated our method is described in section \ref{sc:moti}. The proposed models and estimation procedure are provided in sections \ref{sc:model} and \ref{sc:est} respectively. To evaluate the performance of our method in comparison with existing methods, an extensive simulation study is presented in section \ref{sc:simu}. A real data application of the proposed method to a neuroscience dataset is given in section \ref{sc:application}, followed by a discussion in section \ref{sc:diss} and an appendix. Additional information can be found in the document of supplement materials.

\section{Motivating dataset in neuroscience}\label{sc:moti}
The subjects in the motivating dataset were from our ongoing study at Columbia University Medical Center: the Reference Ability Neural Network (RANN) study.\cite{Stern2009,Stern2014,Stern2018} The RANN study aims to identify neural networks associated with cognitive performance throughout the lifespan of a healthy sample in the four reference abilities (RAs): episodic memory, perceptual speed, fluid reasoning, and vocabulary.\cite{Salthouse2009}

\subsection{Outcome variables}
Based on a previous principal axis factor analysis,\cite{Salthouse2015} a composite score for each RA was obtained using measures from a battery of well-established neuropsychological instruments, and higher scores indicate better performance. The four RA scores for memory, perceptual speed, fluid reasoning, and vocabulary are the outcome variables in our mediation analyses. 

\subsection{Mediators: Imaging measures}
Detailed information on image processing has been reported in a previous study.\cite{moura2019relationship} All images were acquired on the same 3.0 T Philips Achieva Magnet. Structural T1 scans were reconstructed using FreeSurfer v5.1 (\url{http://surfer.nmr.mgh.harvard.edu/}). Each participant’s boundaries of white and gray matter, as well as gray matter and cerebral-spinal-fluid boundaries, were visually inspected slice by slice, and manual control points were added in case of any visible discrepancy. The five lobar regions: frontal, parietal, temporal, cingulate, and occipital, were defined as the lobes, and the volume of WMH in each of the five lobes was automatically extracted. The mediators in our model are the volumes of WMH which are counts.

\subsection{Challenges in exploring mediation effects}
Because older adults commonly have brain WMH,\cite{habes2016white,merino2019white} and WMH is often related to cognitive impairment,\cite{Prins2015} we wish to investigate the mediation effect of these potential mediators, WMH counts in different lobes, between age and RAs. Aging might affect cognitive abilities, but the effects could also occur by being transmitted through WMH in the brain. However, all five potential mediator variables (i.e., WMH) have a large number of zero values present, ranging from 31\% to 76\% (see Table \ref{table_realdata1}) among the 350 subjects. The 6th potential mediator variable, total, is the summation of WMH in five lobes.

\begin{figure}
  \begin{center}
  \includegraphics[width =1\textwidth,angle=0]{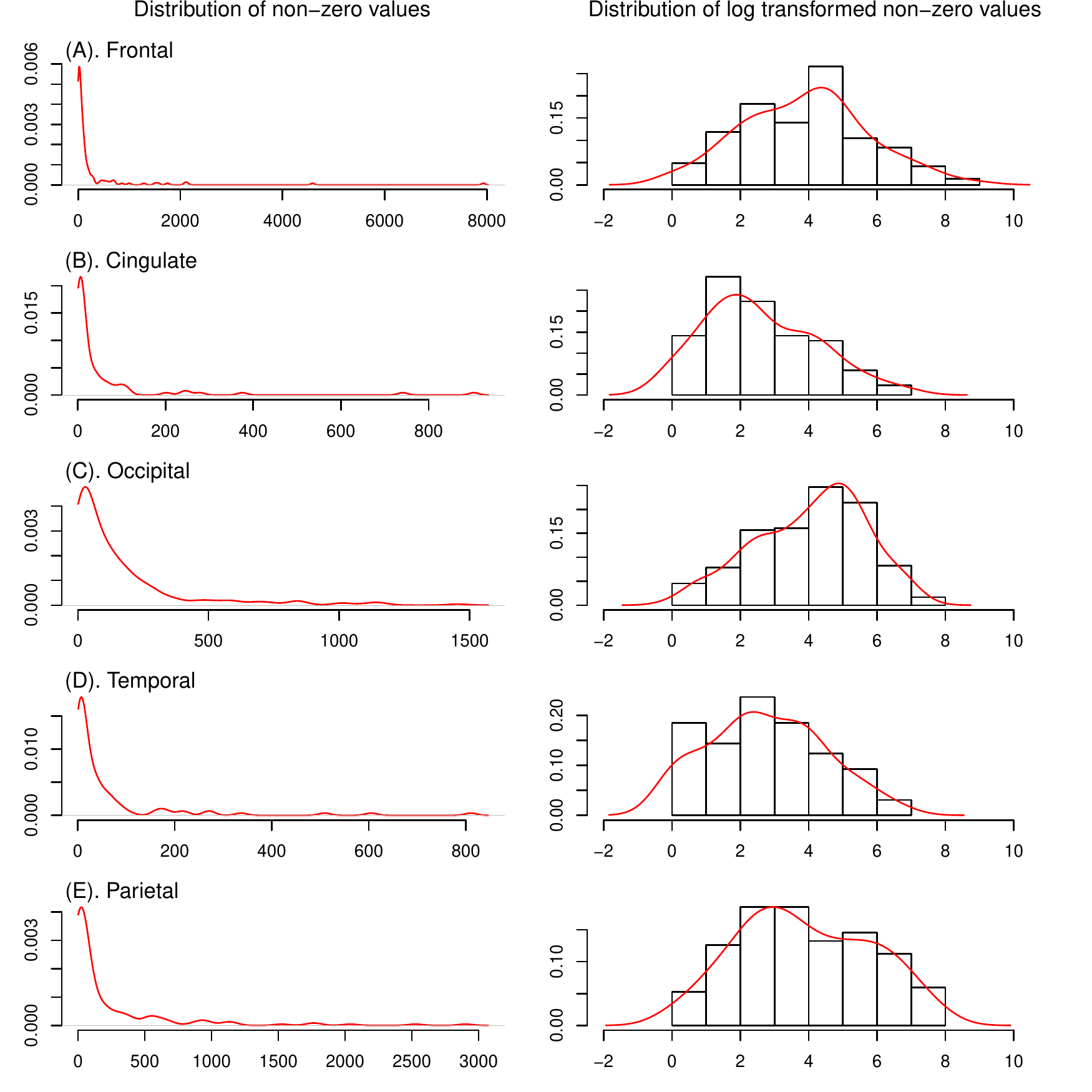}
  \end{center}
  \caption{Distributions of non-zero values (left) and the logarithm transformation (right) for five mediator variables: (A) Frontal, (B) Cingulate, (C) Occipital, (D) Temporal, and (E) Parietal WMH.}
  \label{fig_Mediators}
\end{figure}

\begin{table}
\caption{Counts and proportion of zero measurement values in six potential mediator variables with a sample size of $n=350$.}
\bigskip
\label{table_realdata1}
\centering
\begin{tabular}{c | c c c} \hline
Potential mediator & Zero counts & Non-zero counts & Proportion of zeros\\[1ex] 
\hline\hline
Frontal & 207 & 143 & 0.59\\
Cingulate & 265 & 85 & 0.76 \\
Occipital & 107 & 243 & 0.31\\
Temporal & 253 & 97 & 0.72\\
Parietal & 199 & 151 & 0.57\\
Total & 76 & 274 & 0.22\\[1ex]
\hline \end{tabular}
\end{table}
Moreover, not all zero WMH values indicate these participants were free of underlying pathologies since the lesion counts at the location might be too small to be detected at the time of the MRI scan. In this situation, although the standard mediation analyses could give a rough estimate of the mediation effect, ignoring the large proportion of zero values or even falsely observed zeros make the results less accurate and reliable. Motivated by this deficiency and the common occurrence of such data in practice, we develop a mediation model for zero-inflated mediators. The probability of observing false zeros is also addressed. Due to the high level of skewness of the WMH count data, we assumed that the underlying distribution of the mediator followed a zero-inflated log-normal distribution or a zero-inflated negative binomial distribution. The distributions of the non-zero values before and after taking natural-log transformation are provided in Figure \ref{fig_Mediators}.

\section{Model and Notation}\label{sc:model}
For ease of presentation, the subject index is suppressed throughout this section. Consider an independent variable $X$, a zero-inflated mediator $M$, and a outcome variable $Y$. Note that the binary variable $1_{(M>0)}$ is also considered as a mediator in addition to $M$, where $1_{(\cdot)}$ is an indicator function. We denote $Y_{xm1_{(m>0)}}$ as the potential outcome of $Y$ when $(X, M, 1_{(M>0)})$ take the value of $(x,m,1_{(m>0)})$. Let $M_x$ represent the value of $M$ if $X$ is taking the value of $x$. Let $1_{(M_x>0)}$ denote the value of $1_{(M>0)}$ if $X$ takes the value of $x$.

\subsection{Definitions of mediation effect and direct effect}\label{NIE_NDE}
Following the potential-outcomes formulations,\cite{VanderWeele2009,Imai2010,Pearl2001} the natural indirect effect (NIE), natural direct effects (NDE) and controlled direct effect (CDE) can be defined. The NIE is also called the mediation effect. The total effect of the independent variable $X$ is equal to the summation of NIE and NDE. The average NIE, NDE and CDE if $X$ changes from $x_1$ to $x_2$ are given by:
\begin{align}
&\text{NIE}=E\big(Y_{x_2M_{x_2}1_{(M_{x_2}>0)}}-Y_{x_2M_{x_1}1_{(M_{x_1}>0)}}\big), \label{defNIE}\\
&\text{NDE}=E\big(Y_{x_2 M_{x_1}1_{(M_{x_1}>0)}}-Y_{x_1 M_{x_1}1_{(M_{x_1}>0)}}\big), \label{defNDE}\\
&\text{CDE}=E\big(Y_{x_2 m1_{(m>0)}}-Y_{x_1 m1_{(m>0)}}\big),\hspace{0.1cm}\text{for a fixed (i.e., controlled) value of}\hspace{0.1cm} M, \label{defCDE}
\end{align}

\noindent where $Y_{x_2M_{x_1}1_{(M_{x_1}>0)}}$ is a potential (i.e., counterfactual) outcome. 
Notice $M$ and $1_{(M>0)}$ become two causally ordered, sequential mediators as shown in Figure \ref{fig_MediatPath}. Based on their sequential order, NIE can be further decomposed \cite{Steen2017}: 
\begin{align}
\text{NIE}&=E\big(Y_{x_2M_{x_2}1_{(M_{x_2}>0)}}-Y_{x_2M_{x_1}1_{(M_{x_1}>0)}}\big)\nonumber\\
&=E\big(Y_{x_2M_{x_2}1_{(M_{x_2}>0)}} - Y_{x_2M_{x_1}1_{(M_{x_2}>0)}}\big) + E\big(Y_{x_2M_{x_1}1_{(M_{x_2}>0)}}  - Y_{x_2M_{x_1}1_{(M_{x_1}>0)}}\big) \nonumber\\
&\coloneqq \text{NIE}_1+\text{NIE}_2, \label{decomposeNIE}
\end{align}
where NIE$_1$ is the mediation effect through $M$ summing the two causal pathways $X\rightarrow M\rightarrow Y$ and $X\rightarrow M\rightarrow 1_{(M>0)}\rightarrow Y$, and $\text{NIE}_2$ is the mediation effect through only $1_{(M>0)}$ on the causal pathway $X\rightarrow 1_{(M>0)}\rightarrow Y$.\cite{Steen2017} The NIE, NDE, CDE as well as the decomposition of NIE into summation of $\text{NIE}_1$ and $\text{NIE}_2$ in equation (\ref{decomposeNIE}) are identifiable following the assumptions in subsection \ref{assump}.\cite{Steen2017}

\subsection{Model specifications}
Notice the decomposition in equation (\ref{decomposeNIE}) can be applied to general cases such as continuous, binary, count, and survival outcomes, but we focus on continuous outcomes in this paper. With a continuous outcome variable $Y$, we construct a full mediation model in this subsection. First, the following equation will be used to model the association between $Y$ and $(M,X)$:
\begin{align}
Y_{xm1_{(m>0)}}=Y_{X=x,M=m,1_{(M>0)}=1_{(m>0)}}=\beta_0+\beta_1m+\beta_2 1_{(m>0)}+\beta_3x+\beta_4x1_{(m>0)}+\beta_5xm+\epsilon \label{ymodel}
\end{align}
Equation (\ref{ymodel}) is essentially a regression model allowing for possible interactions between the independent variable $X$ and the binary mediator $1_{(M>0)}$ as well as the continuous mediator $M$ respectively, $\epsilon$ is the random error following the normal distribution $N(0,\delta^2)$, and $\beta_1$, $\beta_2$, $\beta_3$, $\beta_4$ and $\beta_5$ are regression coefficients. Notice that interactions between $X$ and the two mediators $M$ and $1_{(M>0)}$ are accommodated by the product terms $\beta_4X1_{(M>0)}$ and $\beta_5XM$ in the model, which is an advantage of potential-outcomes mediation analysis approaches.\cite{VanderWeele2009,Imai2010} In traditional mediation analysis approaches, the mediation effect can no longer be estimated from the product of coefficients estimates when the interaction between the independent variable and the mediator is present in the regression model.

In addition, for the zero-inflated mediator $M$, a general form for its two-part density or mass function can be written as:
\begin{align}
&f(m;\theta)=\begin{cases}
    \Delta, & m=0\\
    (1-\Delta)G(m;\theta), &m>0
  \end{cases}, \label{zid}
\end{align}
where $0<\Delta<1$ is the probability of $M$ taking value of $0$, $\theta$ is a $K$-dimensional parameter vector associated with the distribution of the positive values conditional on $M$ being positive, and $G(m;\theta)$ is the conditional probability mass or density function of the positive values. To model the dependence of $M$ on $X$, we can allow the parameters $\Delta$ and $\theta$ to depend on $X$. We will construct mediation models using ZILoN, ZINB and ZIP distributions for the mediators. For simplicity, confounders are not included in the equations in this paper, but they can be easily incorporated into the model.

\subsection{Model for zero-inflated log-normal (ZILoN) mediators}\label{zilonSec}
For a ZILoN mediator, the conditional probability density function $G(m;\theta)$ in equation (\ref{zid}) follows a log-normal distribution:
\begin{align}
&G(m;\theta)=\phi(m;\mu,\sigma)=\frac{1}{m\sigma\sqrt{2\pi}}\exp{\Big(-\frac{\big(\log{(m)}-\mu\big)^2}{2\sigma^2}\Big)}, m>0
\end{align}
where the parameter vector $(\mu,\sigma)^T$ corresponds to $\theta$ in the general equation in (\ref{zid}), and $\phi(\cdot)$ is the density function of the log-normal distribution indexed by the parameters $\mu$ and $\sigma$ which are the expected value and standard deviation, respectively, of the random variable after natural-log transformation. 
Equation (\ref{zid}) can be rewritten as: 
\begin{align}
&  f(m;\theta)=\begin{cases}
    \Delta, & m=0\\
    (1-\Delta)\phi(m;\mu,\sigma), &m>0
  \end{cases}, \label{zilon}
\end{align}
The ZILoN mediator $M$ depends on $X$ through the following equations: 
\begin{align}
&\mu=\alpha_{0}+\alpha_{1}X, \label{transzilon1}\\
&\log{\bigg(\frac{\Delta}{1-\Delta}\bigg)}=\gamma_0+\gamma_1X.  \label{transzilon2}
\end{align}
Equations (\ref{ymodel}), (\ref{transzilon1}) and (\ref{transzilon2}) together form the full mediation model for a ZILoN mediator and a continuous outcome. In this paper, the corresponding mediation models constructed for ZINB and ZIP mediators are included in section \ref{ZINBSec} and \ref{ZIPSec} respectively.

\subsection{Model for zero-inflated negative binomial (ZINB) mediators}\label{ZINBSec}
The probability mass function of a negative binomial (NB) distribution for a count random variable $M$ is given by:  
\begin{align}
&  f(m;\mu,r)= \frac{\Gamma(r+m)}{\Gamma(r)m!}\Big(\frac{\mu}{r+\mu}\Big)^m\Big(\frac{r}{r+\mu}\Big)^r, m=0,1,2,\dots
\end{align}
where $\Gamma(\cdot)$ is the gamma function. The random variable $M$ equals the number of failures to achieve $r$ success in a sequence of Bernoulli trials with success probability $p=\frac{r}{r+\mu}$. The parameter $r>0$ can be extended to any positive real value in the above formula. The expected value of $M$ is given by $E(M)=\mu$ and its variance by $Var(M)=\mu+\frac{\mu^2}{r}$, where $r$ is the called the dispersion or shape parameter that controls amount of over-dispersion.  

If the mediator $M$ has a zero-inflated negative binomial distribution, the conditional probability density function $G(m;\theta)$ in equation (\ref{zid}) is the positive part of a NB distribution:
\begin{align}
&G(m;\theta)=\frac{\Gamma(r+m)}{\Gamma(r)m!}\frac{(\frac{\mu}{r+\mu})^m}{(\frac{r}{r+\mu})^{-r}-1},  m=1,2,\dots
\end{align}
The two-part density function in equation (\ref{zid}) for a ZINB mediator $M$ is given by:
\begin{align}
&  f(m;\theta)=\begin{cases}
    \Delta=\Delta^*+(1-\Delta^*)(\frac{r}{r+\mu})^r, & m=0\\\\
    (1-\Delta) \frac{\Gamma(r+m)}{\Gamma(r)m!}\frac{(\frac{\mu}{r+\mu})^m}{(\frac{r}{r+\mu})^{-r}-1}, & m=1,2,\dots
  \end{cases}, \label{zinb}
\end{align}
where the parameter vector $(\mu,r)^T$ corresponds to $\theta$ in equation (\ref{zid}) and controls the number of zeros generated from the NB distribution, $0<\Delta^*<1$ is the parameter controlling the number of excessive zeros (i.e., not generated from the NB distribution), $r$ is the dispersion parameter, and $\mu$ is the expectation of the negative binomial distribution. The ZINB mediator $M$ depends on $X$ through the following equations: 
\begin{align}
&\log(\mu)=\alpha_{0}+\alpha_{1}X, \label{transzinb1}\\
&\log{\bigg(\frac{\Delta^*}{1-\Delta^*}\bigg)}=\gamma_0+\gamma_1X.  \label{transzinb2}
\end{align}
Equations (\ref{ymodel}), (\ref{transzinb1}) and (\ref{transzinb2}) together form the full mediation model for a ZINB mediator and a continuous outcome. 

\subsection{Model for zero-inflated Poisson (ZIP) mediators}\label{ZIPSec}
The conditional probability density function $G(m;\theta)$ in equation (\ref{zid}) is the positive part of a Poisson distribution:
\begin{align}
&G(m;\theta)=\frac{\lambda^m}{m! (\exp{(\lambda)}-1)},  m=1,2,\dots
\end{align}
Equation (\ref{zid}) for a ZIP mediator can be rewritten as: 
\begin{align}
&  f(m;\theta)=\begin{cases}
    \Delta=\Delta^*+(1-\Delta^*)\exp{(-\lambda)}, & m=0\\
    (1-\Delta)\frac{\lambda^m}{m! (\exp{(\lambda)}-1)}, & m=1,2,\dots
  \end{cases}, \label{zip}
\end{align}
where $\lambda>0$ is the mean of the Poisson distribution. $\lambda$ controls the number of zeros generated by the data generating process underlying the Poisson distribution, while $0<\Delta^*<1$ controls the number of excessive zeros in addition to zeros from the Poisson distribution. The ZIP mediator $M$ depends on $X$ through the following equations: 
\begin{align}
&\log{(\lambda)}=\alpha_0+\alpha_1X , \label{transzip1}\\
&\log{\bigg(\frac{\Delta^*}{1-\Delta^*}\bigg)}=\gamma_0+\gamma_1X. \label{transzip2}
\end{align}
Equations (\ref{ymodel}), (\ref{transzip1}) and (\ref{transzip2}) together form the full mediation model for a ZIP mediator and a continuous outcome.

\subsection{Probability mechanism for observing false zeros}\label{pfalse0}
It is common to observe two types of zeros for $M$ in a dataset with excessive zeros: true zeros and false zeros. In our real data example, a true zero represents that a participant is free of lesion in a specific lobe, and a false zero means he does have white matter damage but it is indiscernible given the accuracy of the medical device or image processing pipeline. We use $M$ to denote the true value of the mediator which is true WMH in the example data, and use $M^*$ for the observed value of $M$. When the observed value of the mediator is positive (i.e., $M^*>0$), we assume  $M^*=M$. However, when $M^*=0$, we don't know whether $M$ is truly zero or $M$ is positive but incorrectly observed as zero. We consider the following mechanism for observing a zero: 

\begin{equation}\label{zeroMecha}
  P(M^*=0|M)=\begin{cases}
    \exp(-\eta^2 M), & M\le B\\
    0, &M>B
  \end{cases},
\end{equation}
where the parameter $\eta$ needs to be estimated, and $B>0$ is a known constant. When $M=0$, the observed value will be 0 with probability 1. When $M \le B$, the observed value will be 0 with the probability of $\exp(-\eta^2 M)$, which means the larger the true value, the smaller the chance this value is observed as a zero. On the other hand, when $M>B$, the probability that it is being observed as zero is 0. That is, when the mediator value is large enough (e.g., a large number of lesions in the motivating example), it is unlikely that the measurement is undetectable or falsely observed as zero. Note that the chance of observing false zeros is assumed to depend on true underlying mediator values only but not independent variable $X$ or any confounders $Z$ (i.e., $P(M^*=0|M)=P(M^*=0|M,X,Z)$). The value of $B$ can be informed on the basis of the insights and judgements of professionals in the specific field from which the data arose. Here we set $B = 20$ based on suggestions from experts in brain imaging research.

\subsection{Effects for ZILoN mediators}
\begin{figure}
  \begin{center}
  \includegraphics[width =1\textwidth,angle=0]{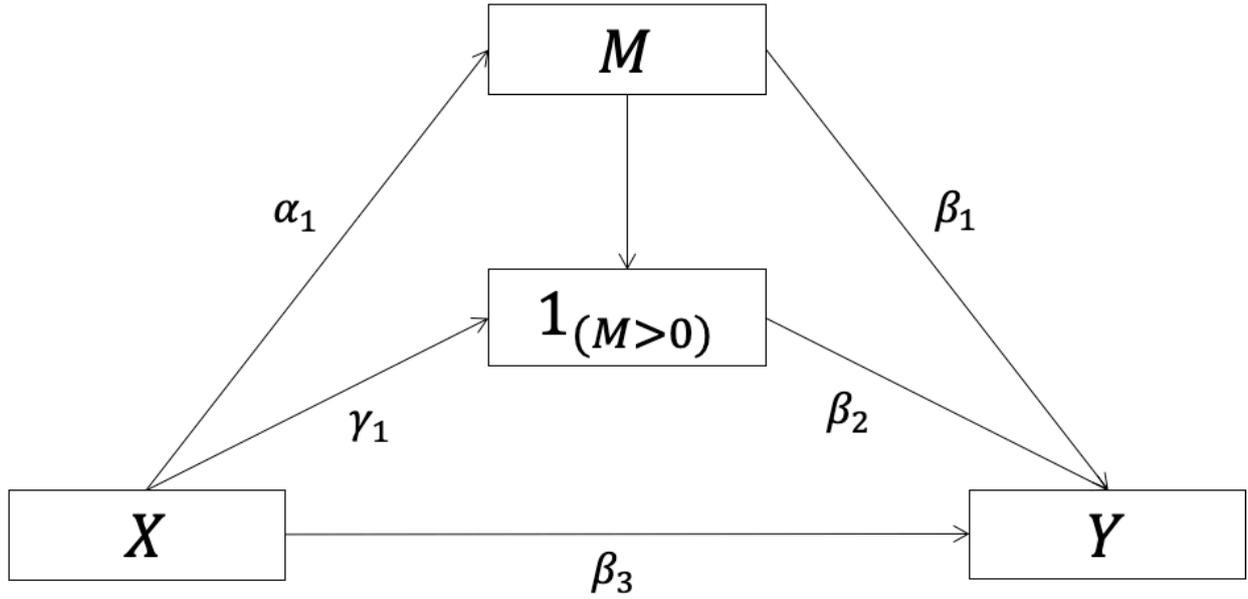}
  \end{center}
  \caption{Potential causal mediation pathways of a zero-inflated mediator.}
  \label{fig_MediatPath}
\end{figure}

The NIE, NDE, and CDE are derived for the proposed mediation model. By plugging the equations (\ref{ymodel}) and (\ref{zilon})-(\ref{transzilon2}) into the definitions of effects in subsection \ref{NIE_NDE} and using Riemann-Stieltjes integration,\cite{TerHorst1986} we obtain the following expressions for ZILoN mediators:
\begin{align*}
\text{NIE}_1&=E\big(Y_{x_2M_{x_2}1_{(M_{x_2}>0)}} - Y_{x_2M_{x_1}1_{(M_{x_2}>0)}}\big) \\
&=E\Big[E(Y_{x_2 M_{x_2}1_{(M_{x_2}>0)}}|M_{x_2},1_{(M_{x_2}>0)})\Big]-E\Big[E(Y_{x_2 M_{x_1}1_{(M_{x_2}>0)}}|M_{x_1},1_{(M_{x_2}>0)})\Big] \\ 
&=E\Big(\beta_0+\beta_1M_{x_2}+\beta_2 1_{(M_{x_2}>0)}+\beta_3x_2+\beta_4x_21_{(M_{x_2}>0)}+\beta_5x_2M_{x_2}\Big) \\
&\hspace{0.5cm}-E\Big(\beta_0+\beta_1M_{x_1}+\beta_2 1_{(M_{x_2}>0)}+\beta_3x_2+\beta_4x_21_{(M_{x_2}>0)}+\beta_5x_2M_{x_1}\Big) \\
&=(\beta_1+\beta_5x_2)(E(M_{x_2})-E(M_{x_1})) \\
&=(\beta_1+\beta_5x_2)\Bigg[\int\limits_{m\in\Omega}mdF_{M_{x_2}}(m)-\int\limits_{m\in\Omega}mdF_{M_{x_1}}(m)\Bigg] \\
&=(\beta_1+\beta_5x_2)\Bigg[(1-\Delta_{x_2})\int\limits_{m\in\Omega\setminus 0}m\phi(m;\mu_{x_2},\sigma)dm-(1-\Delta_{x_1})\int\limits_{m\in\Omega\setminus 0}m\phi(m;\mu_{x_1},\sigma)dm\Bigg] \\
&=(\beta_1+\beta_5x_2)\Bigg[(1-\Delta_{x_2})\exp{\Big(\mu_{x_2}+\frac{\sigma^2}{2}\Big)}-(1-\Delta_{x_1})\exp{\Big(\mu_{x_1}+\frac{\sigma^2}{2}\Big)}\Bigg], \\
\text{NIE}_2&=E\big(Y_{x_2M_{x_1}1_{(M_{x_2}>0)}}  - Y_{x_2M_{x_1}1_{(M_{x_1}>0)}}\big) \\
&=E\Big[E(Y_{x_2 M_{x_1}1_{(M_{x_2}>0)}}|M_{x_1},1_{(M_{x_2}>0)})\Big]-E\Big[E(Y_{x_2 M_{x_1}1_{(M_{x_1}>0)}}|M_{x_1},1_{(M_{x_1}>0)})\Big] \\ 
&=E\Big(\beta_0+\beta_1M_{x_1}+\beta_2 1_{(M_{x_2}>0)}+\beta_3x_2+\beta_4x_21_{(M_{x_2}>0)}+\beta_5x_2M_{x_1}\Big) \\
&\hspace{0.5cm}-E\Big(\beta_0+\beta_1M_{x_1}+\beta_2 1_{(M_{x_1}>0)}+\beta_3x_2+\beta_4x_21_{(M_{x_1}>0)}+\beta_5x_2M_{x_1}\Big) \\
&=(\beta_2+\beta_4x_2)(E(1_{(M_{x_2}>0)})-E(1_{(M_{x_1}>0)})) \\
&=(\beta_2+\beta_4x_2)(\Delta_{x_1}-\Delta_{x_2}), \\
\text{NDE}&=E\big(Y_{x_2 M_{x_1}1_{(M_{x_1}>0)}}-Y_{x_1 M_{x_1}1_{(M_{x_1}>0)}}\big) \\
&=\beta_3(x_2-x_1)+\beta_4(x_2-x_1)(1-\Delta_{x_1})+\beta_5(x_2-x_1)(1-\Delta_{x_1})\exp{\Big(\mu_{x_1}+\frac{\sigma^2}{2}\Big)} \\
&=(x_2-x_1)\Bigg\{\beta_3+(1-\Delta_{x_1})\Big[\beta_4+\beta_5\exp{\Big(\mu_{x_1}+\frac{\sigma^2}{2}\Big)\Big]}\Bigg\}, \\
\text{CDE}&=E\big(Y_{x_2 m1_{(m>0)}}-Y_{x_1 m1_{(m>0)}}\big) \\
&=\beta_3(x_2-x_1)+\beta_4(x_2-x_1)1_{(m>0)}+\beta_5(x_2-x_1)m \\
&=(x_2-x_1)\big(\beta_3+\beta_41_{(m>0)}+\beta_5m\big),  
\end{align*}
where $\Omega$ denotes the domain of the mediator $M$, $\Omega\setminus 0$ denotes the subset of $\Omega$ that does not contain $0$, $F_{M_x}(m)$ denotes the cumulative distribution function of $M_x$, and $dF_{M_x}(m)$ denotes the Stieltjes integration \cite{TerHorst1986} with respect to $F_{M_x}(m)$. Figure \ref{fig_MediatPath} depicts the directed acyclic graph of the proposed method. $\text{NIE}_1$ can be interpreted as the mediation effect through $M$ summing the two causal pathways $X\rightarrow M\rightarrow Y$ and $X\rightarrow M\rightarrow 1_{(M>0)}\rightarrow Y$, and $\text{NIE}_2$ is the mediation effect through only $1_{(M>0)}$ on the causal pathway $X\rightarrow 1_{(M>0)}\rightarrow Y$.\cite{Steen2017} This decomposition can also be seen in Figure \ref{fig_MediatPath} where there are three possible indirect causal pathways from $X$ to $Y$: two causal pathways through the mediator $M$ (corresponds to $\text{NIE}_1$) and one causal pathway through only the mediator $1_{(M>0)}$ (corresponds to $\text{NIE}_2$).
The NIE, $\text{NIE}_1$, $\text{NIE}_2$, NDE and CDE can be estimated by plugging the estimates of model parameters into the above expressions. In this paper, confidence intervals (CI) for NIE are obtained using the multivariate delta method. An alternative approach to construct CI is bootstrapping.\cite{Efron1986} When the distribution of mediator is not zero-inflated, $\text{NIE}_2$ becomes 0 since $\Delta_x$ reduces to 0, and thus the NIE reduces to a usual NIE that can be calculated by standard approaches.\cite{VanderWeele2009, Imai2010} Formulas for ZINB and ZIP mediators can be found in the supplementary materials.

\subsection{Assumptions}\label{assump}
To draw causal inference for counterfactual-outcomes mediation analyses with the two sequential mediators $M$ and $1_{(M>0)}$ as shown in Figure \ref{fig_MediatPath}, some assumptions are needed \cite{Steen2017} to identify terms involving counterfactual variables. The first four assumptions are: 
\begin{align}
&Y_{x'm1_{(m>0)}}\ind X | Z,\label{assump1}\\
&Y_{x'm1_{(m>0)}}\ind \{M_x, 1_{(M_x>0)}\} | X=x,Z,\label{assump2}\\
&\{M_x, 1_{(M_x>0)}\}\ind X | Z,\label{assump3}\\
&Y_{x'm1_{(m>0)}}\ind \{M_x, 1_{(M_x>0)}\} | Z,\label{assump4}
\end{align}
where $Z$ denotes observed confounders (if any), and  the notation $U \ind V | W$ means that random variables $U$ and $V$ are conditionally independent given $W$ which could be a vector. These assumptions are the same as those for the MSM method,\cite{VanderWeele2009,multipleMedi14} and they have straightforward interpretations. The first three assumptions in equations (\ref{assump1}) - (\ref{assump3}) mean that there are no unmeasured confounding for $X-Y$, $\{M,1_{(M>0)}\}-Y$ and $X-\{M,1_{(M>0)}\}$ associations. The fourth assumption in equation (\ref{assump4}) requires that there are no post-baseline confounders of the $\{M,1_{(M>0)}\}-Y$ association affected by $X$. As Steen et al \cite{Steen2017} stated, two additional assumptions are required regarding the causal effect of the first mediator on the second mediator when the two mediators are sequential: (a) The effect of the first mediator on the second mediator is not confounded within strata of $\{X, Z\}$ and (b) None of the confounders (if any) for the effect of the first mediator on the second mediator is affected by X. In our setting, these two additional assumptions are automatically satisfied because $1_{(M>0)}$ is a deterministic function of $M$. 

Notice that the mediator $1_{(M>0)}$ can be dropped from the first four assumptions (\ref{assump1}) - (\ref{assump4}) because it is a deterministic function of $M$. In other words, it is sufficient to have the first four assumptions satisfied for the mediator $M$, and that implies the first four assumptions satisfied for $1_{(M>0)}$. We included $1_{(M>0)}$ in the first four assumptions regardless to make the coverage of the assumptions more explicit. It is also worth noting that there are causal mediation analysis approaches in the literature that can relax some of the assumptions on unmeasured confounding.\cite{Zheng2014,Guo2018}

\section{Estimation}\label{sc:est}
\subsection{Estimation with EM algorithm}\label{generalEM}
We construct the log-likelihood function for the zero-inflated mediator model and implement maximum likelihood estimation (MLE) by employing the Expectation Maximization (EM) algorithm to estimate model parameters and mediation effects. Some details of the derivations for the log-likelihood function and the EM algorithm are provided in section \ref{sc:appendix_a} of the Appendix. 

\subsubsection{Log-likelihood function}
An indicator variable $C=1_{(M\ne0)}$ is defined such that $C_i=0$ if the $i$th subject has the true value of $M$ equal to 0 and $C_i=1$ if the $i$th subject has a non-zero $M$. Notice that $C_i$ is a latent variable for those subjects having false zeros. Let $R=1_{(M^*\ne0)}$ denote the observed indicator variable. Based on the mechanism of observing false zeros in section \ref{pfalse0}, we know that $C_i=1$ for individuals with $R_i=1$. For individuals with $R_i=0$, it is not known whether they have true zero values ($C_i=0$) or if they were incorrectly observed as false zeros ($C_i=1$). Thus, $C_i$ will be treated as missing data for those subjects with $R_i=0$ in the EM algorithm procedure. The observed data variable vector is $(y_i,m_i^*,r_i,c_i,x_i)$ for the $i$th subject, and the log-likelihood function is:
\begin{align*}
\ell&=\log\Big(\prod_{i=1}^Nf(y_i,m_i^*,r_i,c_i|x_i)\Big)\\
&=\sum_{i=1}^N\Bigg\{1_{(c_i=0)}\Big[\log(\Delta_i)+\log(f(y_i,m_i^*,r_i|x_i,c_i=0))\Big]+1_{(c_i=1)}\Big[\log(1-\Delta_i)+\log(f(y_i,m_i^*,r_i|x_i,c_i=1))\Big]\Bigg\},
\end{align*}
where $P(C_i=0)=P(M_i=0)=\Delta_i$ is the probability of having a true zero mediator value, $f(y_i,m_i^*,r_i,c_i|x_i)$ is the joint density function conditional $x_i$, and $f(y_i,m_i^*,r_i|x_i,c_i)$ is the joint density function conditional $x_i$ and $c_i$. Let $\ell_{i0}=\log{\big(f(y_i,m_i^*,r_i|x_i,c_i=0)\big)}$ and $\ell_{i1}=\log{\big(f(y_i,m_i^*,r_i|x_i,c_i=1)\big)}$. Then
\begin{align*}
&\ell=\sum_{i=1}^N\Bigg\{1_{(c_i=0)}\Big[\log(\Delta_i)+\ell_{i0}\Big]+1_{(c_i=1)}\Big[\log(1-\Delta_i)+\ell_{i1}\Big]\Bigg\}.
\end{align*}
We divide all subjects into two groups by whether $M_i^*$ is zero. The first group consists of subjects whose $M_i^*$ are positive (i.e., $R_i=C_i=1$ and $M_i^*>0$). The second group consists of subjects whose $M_i^*$ are zeros (i.e., $R_i=0$ and $M_i^*=0$, but $C_i$ and $M_i$ are unobserved). Furthermore, group 2 may have two subgroups: some of them have true zeros (i.e., $C_i=0$, $M_i=0$), while the rest have false zeros (i.e, $C_i=1$, $M_i>0$). The complete log-likelihood function can be written as:
\begin{align}\label{completell}
\ell&=\sum_{i\in\text{group1}}\Big[\log(1-\Delta_i)+\ell_{i1}^1\Big]+\sum_{i\in\text{group2}}\Bigg\{1_{(c_i=0)}\Big[\log(\Delta_i)+\ell_{i0}^2\Big]+1_{(c_i=1)}\Big[\log(1-\Delta_i)+\ell_{i1}^2\Big]\Bigg\},
\end{align}
where $\ell_{i1}^1$ denotes $\ell_{i1}$ in group 1, and $\ell_{i0}^2$ and $\ell_{i1}^2$ denote $\ell_{i0}$ and $\ell_{i1}$ in group 2 respectively. The further derivations of terms involved are specific to the distribution of zero-inflated mediators (see subsection \ref{sc:loglike_zilon}).

So far, all the terms in the complete log-likelihood function $\ell$ in equation (\ref{completell}) can be obtained except that $1_{(c_i=0)}$ and $1_{(c_i=1)}$ are not observable (i.e., missing data) for group 2, and thus the EM algorithm is used for estimation. 

\subsubsection{Expectation step (E step)}
Let $\Theta$ denote the vector including all parameters and $\Theta^0$ denote the initial value of the parameter vector $\Theta$.
Denote $Q(\Theta|\Theta^0)$ as the expectation of the log-likelihood function with respect to the conditional distribution of the latent variable, $C_i$, given observed data and the current parameter value (i.e., $\Theta^0$):
\begin{align*}
Q(\Theta|\Theta^0)&=E_{c_i|y_i,m_i^*,r_i,x_i,\Theta^0}(\ell)\nonumber\\
&=\sum_{i\in\text{group1}}\Big[\log(1-\Delta_i)+\ell_{i1}^1\Big]\nonumber\\
&\hspace{0.4cm}+\sum_{i\in\text{group2}}\Bigg(E_{c_i|y_i,m_i^*,r_i,x_i,\Theta^0}\big(1_{(c_i=0)}\big)\Big[\log(\Delta_i)+\ell_{i0}^2\Big]+E_{c_i|y_i,m_i^*,r_i,x_i,\Theta^0}\big(1_{(c_i=1)}\big)\Big[\log(1-\Delta_i)+\ell_{i1}^2\Big]\Bigg).
\end{align*}
Let $\tau_{ik}(\Theta^0)=E_{c_i|y_i,m_i^*,r_i,x_i,\Theta^0}\big(1_{(c_i=k)}\big)$ for $k=0,1$. By using the Bayes formula, we have
\begin{align*}
\tau_{i0}(\Theta^0)&=E\big(1_{(c_i=0)}|y_i,m_i^*=0,r_i=0,x_i,\Theta^0\big)=\frac{\Delta_i\exp{(\ell_{i0}^2)}}{\Delta_i\exp{(\ell_{i0}^2)}+(1-\Delta_i)\exp{(\ell_{i1}^2)}}\Bigg|_{\hspace{0.1cm}\text{evaluated at}\hspace{0.1cm} \Theta^0},\\
\tau_{i1}(\Theta^0)&=E\big(1_{(c_i=1)}|y_i,m_i^*=0,r_i=0,x_i,\Theta^0\big)=1-\tau_{i0}(\Theta^0).
\end{align*}
Finally we get 
\begin{align}\label{logL.Estep2}
Q(\Theta|\Theta^0)&=E(\ell|y_i,m_i^*,r_i,x_i,\Theta^0)\nonumber\\
&=\sum_{i\in\text{group1}}\Big[\log(1-\Delta_i)+\ell_{i1}^1\Big]+\sum_{i\in\text{group2}}\Bigg\{\tau_{i0}(\Theta^0)\Big[\log(\Delta_i)+\ell_{i0}^2\Big]+\tau_{i1}(\Theta^0)\Big[\log(1-\Delta_i)+\ell_{i1}^2\Big]\Bigg\}.
\end{align}

\subsubsection{Maximization step (M step)}
We maximize $Q(\Theta|\Theta^0)$ in equation (\ref{logL.Estep2}) with respect to $\Theta$, and find the maximizer $\Theta^1$:  
\begin{align*}
\Theta^1=\underset{\Theta}{\mathrm{arg\;max}}Q(\Theta|\Theta^0).
\end{align*}
We repeat the E step and M step by replacing $\Theta^0$ with $\Theta^1$ until convergence. The final value of $\Theta^1$ is the MLE. The standard error and CI of the estimator of each parameter are then calculated using the observed Fisher information matrix.\cite{Oakes1999}
The MLEs of direct and mediation effects can be easily derived by plugging the MLEs of parameters into their formulas, and the corresponding standard errors will be determined by the multivariate delta method. 

\subsection{Log-likelihood function for ZILoN mediators}\label{sc:loglike_zilon}
Notice that we need to have $\Delta_i$, $\ell_{i1}^1$, $\ell_{i0}^2$, and $\ell_{i1}^2$ in order to construct the log-likelihood function as in equation (\ref{completell}). Next we derive the expressions of those terms for a ZILoN mediator model (see section \ref{zilonSec}), and detailed derivations can be found in section \ref{sc:appendix_b} of the Appendix. The term $\Delta_i$ is equal to $\exp(\gamma_0+\gamma_1x_i)/\big(1+\exp(\gamma_0+\gamma_1x_i)\big)$ according to equation (\ref{transzilon2}).
For group 1 consisting of subjects whose observed values for the mediator are positive (i.e., $r_i=1$), the log-likelihood contribution from the $i$th individual can be calculated as:
\begin{align*}
\ell_{i1}^1&=\log{\big(f(y_i,m_i^*,r_i=1|x_i,c_i=1)\big)}\\
&=-\log(\delta)-\frac{(y_i-\beta_0-\beta_1m_i^*-\beta_2-(\beta_3+\beta_4)x_i-\beta_5x_im_i^*)^2}{2\delta^2}+\log\Big[1-1_{(m_i^*\le B)}\exp{(-\eta^2m_i^*)}\Big]\\
&\hspace{0.4cm}-\log(m_i^*\sigma)-\frac{(\log (m_i^*)-\mu_i)^2}{2\sigma^2}-\log{(2\pi)},
\end{align*}
where $f(y_i|m_i^*,x_i,c_i)$ is the conditional density function of $Y_i$, $P(R_i=1|m_i^*,x_i,c_i)$ is the conditional probability mass function of $R_i$ at $r_i=1$ and $f_M(m_i^*|x_i,c_i)$ is the conditional probability density/mass function of $M_i$. 

For group 2 subjects whose observed values of the mediator are zeros (i.e., $r_i=0$), we can also derive $\ell_{i0}^2$, the log-likelihood contribution of those individuals with a true zero ($m_i^*=m_i=0$), as follows:
\begin{align*}
\ell_{i0}^2 &= \log(f(y_i,m_i^*=0,r_i=0|x_i,c_i=0))\\
&=-\log(\delta)-\frac{(y_i-\beta_0-\beta_3x_i)^2}{2\delta^2}-0.5\log{(2\pi)}.
\end{align*}

The remainder of the group 2 subjects are individuals with a false zero ($m_i>0$ and $m_i^*=0$). It is more computationally demanding to calculate $\ell_{i1}^2$ for individuals who had false zeros (i.e., $m_i>0$ and $m_i^*=0$). Because we do not know the true value of the false zeros, we average over all possible non-zero values that are less than or equal to the bound $B$ through an integration. The resulting log-likelihood contribution is
\begin{align*}
\ell_{i1}^2&=\log(f(y_i,m_i^*=0,r_i=0|x_i,c_i=1))\\
&=\log\Bigg(\frac{1}{2\pi\delta\sigma}\int\limits_0^B h_i(m)dm\Bigg),
\end{align*}
where 
\begin{align*}
h_i(m)=\frac{1}{m}\exp\bigg(-\frac{(\log (m)-\mu_i)^2}{2\sigma^2}-\frac{(y_i-\beta_0-\beta_1m-\beta_2-(\beta_3+\beta_4)x_i-\beta_5x_im)^2}{2\delta^2}-\eta^2m\bigg).
\end{align*}
With the above derived terms for $\Delta_i$, $\ell_{i1}^1$, $\ell_{i0}^2$, and $\ell_{i1}^2$, an EM algorithm can be used to obtain parameter estimates for ZILoN mediators as illustrated in Section \ref{generalEM}. Derivations for models with ZINB and ZIP mediators are provided in section S1 of the supplementary materials.

\section{Simulation}\label{sc:simu}
Extensive simulations were carried out to demonstrate the performance of the proposed method in comparison with a standard potential-outcomes mediation method, the MSM method.\cite{VanderWeele2009} The IKT approach \cite{Imai2010} had a similar performance to MSM and thus it was not included in the comparison. We outlined 3 models for ZILoN, ZINB, and ZIP mediators respectively, but the underlying distribution of mediators is unknown in practice and a model selection is needed. For the proposed method, the ZILoN, ZINB, and ZIP models were fitted for the mediators, and a final mediation model among the three was chosen by the Akaike information criterion (AIC).

In the following 3 subsections, data were generated under ZILoN, ZINB, and ZIP settings respectively, and the proposed method was applied with the AIC model selection criterion to evaluate the performance as well as robustness for potential model misspecification. A total of 100 random samples were generated with a sample size of 1000 and the outcome $Y$ was randomly generated using equation (\ref{ymodel}) with $\beta_5=0$. To mimic the real data, we considered scenarios with the proportion of all zeros for the mediator ranging from approximately 30\% to 76\% of which about half were false zeros. The value for $B$ was set to be 20 for the probability mechanism of observing false zeros throughout the simulation settings. The R package ``CMAverse'' \cite{Shi2021} was used to implement the MSM method.

We also compared the performances under the scenarios where the mediators are not zero-inflated (i.e., no excessive or false zeros) for ZINB and ZIP mediators in section S2 of the supplementary materials and it showed equivalent performance between the proposed method and MSM although MSM has higher biases. ZILoN mediators were not included in this comparison because ZILoN distributions do not generate zeros when there are no excessive or false zeros.

As suggested by a referee, we investigated the empirical approach that models the two mediators, $M$ and $1_{(M>0)}$, separately which was implemented by the R package ``CMAverse''.\cite{Shi2021} Models for the empirical approach are provided in section S3 of the supplementary materials along with its simulation performance. When the mediator is zero-inflated, it had a poor performance (see Tables S3 - S5) which is likely because (a) the empirical method does not address false zeros, (b) it mis-specifies the model for the mediator due to excessive and false zeros and (c) it also mis-specifies the outcome model with only one mediator. Moreover, when modeling $M$ as the single mediator in the empirical approach, the path $X\rightarrow M\rightarrow 1_{(M>0)}\rightarrow Y$ is missing, which is part of the definition for $\text{NIE}_1$. When the mediator is not zero-inflated (i.e., no excessive or false zeros), the performance (see Tables S6 - S7) remains suboptimal although the results for the binary mediator $1_{(M>0)}$ were substantially improved. 

\subsection{Data with ZILoN mediators}
The ZILoN data were generated with parameter values estimated from the real neuroscience dataset in section \ref{sc:moti} using scaled National Adult Reading Test IQ scores as the outcome variable, age as the independent variable, and Occipital WMH as the mediator. The independent variable, age, used in the simulation was generated from its empirical distribution in the real data. The average mediation effects were calculated for age increasing from 50 years old to 70 years old. The results (see Table \ref{table_simzilon}) showed good performance of the proposed method in terms of bias and the coverage probability (CP) of the 95\% CI across all scenarios where the CP was close to the nominal level 95\%. In all random samples, AIC selected the ZILoN model. 

We did not compare the proposed approach with the MSM in this setting because the R package ``CMAverse'' \cite{Shi2021} does not have an appropriate model for ZILoN mediators due to the zeros. As far as we know, there are no existing R packages applicable to a ZILoN mediator for MSM.

The results of the empirical method (suggested by a referee) that models the single mediator $1_{(M>0)}$ are provided in Table S3 in the supplementary materials. The R package``CMAverse'' was used to implement the MSM method for the empirical method, and the reason that $M$ was not considered in this analysis is because the R package is not applicable to ZILoN mediators as mentioned earlier. The results showed large biases and low CP (see Table S3) and that is likely because (a) the model for $Y$ was misspecified because only one mediator is included and (b) the excessive zeros and false zeros cannot be accounted for by the empirical method. 

\begin{table}
\caption{Simulation results$^+$ for ZILoN data when $X$ changes from $x_1=50$ to $x_2=70$ from the final model chosen by AIC. Mean SE is the mean of estimated standard errors. Bias is the mean estimates minus true values, and percent of bias is the bias as a percentage of true values. CP denotes the empirical coverage probability of the 95\% CI.} 
\label{table_simzilon}
\centering
\resizebox{14cm}{!}{
\begin{tabular}{c c c c c c c c c} \\
\hline
Total zeros (\%)& Method & Mediation & True & Mean & Mean & Bias & Percent of Bias & CP\\
 about half are false zeros& & Effect && Estimate & SE && (\%) & (\%)\\
\hline\hline\\
\multirow{1}{3em}{$\sim 30$} & Proposed & $\text{NIE}_1$ & -0.04 & -0.04 & 0.01 & 0.00 & 4.88 & 96.00 \\ 
&&$\text{NIE}_2$ & 0.80 & 0.76 & 0.12 & -0.04 & -5.01 & 92.00 \\ 
&&NIE & 0.76 & 0.72 & 0.12 & -0.04 & -5.54 & 92.00 \\ 
\hline \\
\multirow{1}{3em}{$\sim 50$} & Proposed & $\text{NIE}_1$ & -0.04 & -0.04 & 0.01 & 0.00 & 2.44 & 95.00 \\ 
&&$\text{NIE}_2$ & 1.11 & 1.07 & 0.13 & -0.04 & -3.34 & 94.00 \\ 
&&NIE & 1.07 & 1.03 & 0.13 & -0.04 & -3.47 & 95.00 \\ 
\hline\\
\multirow{1}{3em}{$\sim 60$} & Proposed & $\text{NIE}_1$  & -0.04 & -0.04 & 0.01 & 0.00 & 2.44 & 97.00 \\
&&$\text{NIE}_2$ & 1.21 & 1.24 & 0.14 & 0.02 & 2.06 & 94.00 \\ 
&&NIE & 1.17 & 1.20 & 0.14 & 0.02 & 2.05 & 94.00 \\ 
\hline\\
\multirow{1}{3em}{$\sim 70$} & Proposed & $\text{NIE}_1$ & -0.04 & -0.04 & 0.01 & 0.00 & 0.00 & 95.00 \\
&&$\text{NIE}_2$ & 1.31 & 1.34 & 0.11 & 0.04 & 2.83 & 97.00 \\ 
&&NIE & 1.27 & 1.31 & 0.11 & 0.04 & 2.83 & 97.00 \\ 
\hline\\
\multirow{1}{3em}{$\sim 76$} & Proposed & $\text{NIE}_1$ & -0.03 & -0.03 & 0.01 & 0.00 & 6.90 & 98.00 \\ 
&&$\text{NIE}_2$ & 1.15 & 1.19 & 0.09 & 0.04 & 3.46 & 93.00 \\ 
&&NIE & 1.13 & 1.16 & 0.09 & 0.04 & 3.37 & 93.00 \\  [1ex]
\hline
\multicolumn{9}{l}{Proposed: proposed method for ZILoN, ZINB, and ZIP mediators chosen by AIC}\\
\multicolumn{9}{l}{$\text{NIE}_1$: mediation effect attributable to the change in the mediator on the numerical scale}\\
\multicolumn{9}{l}{$\text{NIE}_2$: mediation effect attributable to the binary change of the mediator from zero to a non-zero status}\\
\multicolumn{9}{l}{NIE: total mediation effect}\\
\multicolumn{9}{l}{$^+$Results for the MSM method were not included as the existing R package is not applicable}\\[1ex]
\end{tabular}
}
\end{table}

\subsection{Data with ZINB mediators}
The independent variable $X$ in the ZINB data was generated from a standard normal distribution. $\text{NIE}_1$, $\text{NIE}_2$, and NIE were calculated for $X$ increasing from 0 to 1. The simulation results (see Table \ref{table_simzinb}) showed that the proposed model had good performance in terms of bias and CP across all scenarios. For the proposed method, AIC selected the ZINB model for over 94\% of the random samples and selected the ZILoN model for the rest. The MSM method with a NB mediator $M$ and a binary mediator $1_{(M>0)}$ generated large biases and low CP due to its inability to handle false zeros. MSM cannot decompose NIE into NIE$_1$ and NIE$_2$. 

When the distribution of mediators is not zero-inflated (i.e., no excessive or false zeros), all zeros were generated from the NB distribution. The proposed method can still be applied and had slightly better performance than MSM in terms of biases although MSM's performance was reasonably well (see Table S1 in the supplementary materials).

The results for the empirical approach are presented in Tables S4 and S6 in the supplementary materials. When the mediator is zero-inflated, the results (see Table S4) showed large biases and low CP in all scenarios with different percentages of zeros. As mentioned earlier, this poor performance is likely due to mis-specified models for $Y$ and $M$ as well as the missing path from $X\rightarrow M\rightarrow1_{(M>0)}\rightarrow Y$ when only one mediator is modeled. When the mediator is not zero-inflated (i.e., no excessive or false zeros), the results (see Table S6) was still poor for $M$, but the results for $1_{(M>0)}$ improved substantially which may be because it does not suffer from the mis-specified $M$ model anymore. 

\begin{table}
\caption{Simulation results for ZINB data when $X$ changes from $x_1=0$ to $x_2=1$ from the final model chosen by AIC versus MSM. Mean SE is the mean of estimated standard errors. Bias is the mean estimates minus true values, and percent of bias is the bias as a percentage of true values. CP denotes the empirical coverage probability of the 95\% CI.} 
\label{table_simzinb}
\centering
\resizebox{14cm}{!}{
\begin{tabular}{c c c c c c c c c} \\
\hline
Total zeros (\%)& Method & Mediation & True & Mean & Mean & Bias & Percent of Bias & CP\\
about half are false zeros& & Effect && Estimate & SE && (\%) & (\%)\\
\hline\hline\\
\multirow{1}{3em}{$\sim 30$} & Proposed & $\text{NIE}_1$ & -0.03 & -0.03 & 0.03 & 0.00 & -3.12 & 94.00 \\ 
&&$\text{NIE}_2$  & -0.37 & -0.35 & 0.17 & 0.03 & -6.95 & 95.00 \\ 
&&NIE  & -0.41 & -0.38 & 0.16 & 0.03 & -6.90 & 94.00 \\  \\
&MSM&NIE & -0.41 & -0.13 & 0.14 & 0.28 & -68.47 & 42.00 \\ 
\hline \\
\multirow{1}{3em}{$\sim 50$} & Proposed & $\text{NIE}_1$ & 0.04 & 0.04 & 0.04 & 0.00 & -2.56 & 91.00 \\ 
&&$\text{NIE}_2$  & -0.67 & -0.64 & 0.25 & 0.03 & -4.48 & 98.00 \\ 
&&NIE  & -0.63 & -0.60 & 0.23 & 0.03 & -4.75 & 98.00 \\ \\
&MSM&NIE & -0.63 & -0.17 & 0.14 & 0.47 & -73.85 & 7.00 \\ 
\hline\\
\multirow{1}{3em}{$\sim 60$} & Proposed & $\text{NIE}_1$  & 0.06 & 0.06 & 0.05 & 0.00 & -1.75 & 93.00 \\ 
&&$\text{NIE}_2$   & -0.76 & -0.73 & 0.30 & 0.04 & -4.71 & 91.00 \\ 
&&NIE & -0.71 & -0.67 & 0.27 & 0.04 & -5.09 & 91.00 \\   \\
&MSM&NIE & -0.71 & -0.13 & 0.13 & 0.57 & -81.05 & 0.00 \\ 
\hline\\
\multirow{1}{3em}{$\sim 70$} & Proposed & $\text{NIE}_1$ & 0.07 & 0.07 & 0.06 & 0.00 & 5.71 & 90.00 \\ 
&&$\text{NIE}_2$  & -0.82 & -0.80 & 0.35 & 0.02 & -2.20 & 94.00 \\ 
&&NIE  & -0.75 & -0.73 & 0.31 & 0.02 & -3.20 & 94.00 \\  \\
&MSM&NIE & -0.75 & -0.12 & 0.12 & 0.63 & -83.49 & 0.00 \\ 
\hline\\
\multirow{1}{3em}{$\sim 76$} & Proposed & $\text{NIE}_1$ & 0.08 & 0.08 & 0.08 & 0.00 & 5.26 & 91.00 \\ 
&&$\text{NIE}_2$ & -0.85 & -0.82 & 0.53 & 0.02 & -2.60 & 91.00 \\ 
&&NIE  & -0.77 & -0.74 & 0.47 & 0.03 & -3.37 & 91.00 \\   \\
&MSM&NIE  & -0.77 & -0.12 & 0.11 & 0.65 & -84.82 & 0.00 \\  [1ex]
\hline
\multicolumn{9}{l}{Proposed: proposed method for ZILoN, ZINB, and ZIP mediators chosen by AIC}\\
\multicolumn{9}{l}{MSM: standard mediation analysis approach using marginal structural models}\\
\multicolumn{9}{l}{$\text{NIE}_1$: mediation effect attributable to the change in the mediator on the numerical scale}\\
\multicolumn{9}{l}{$\text{NIE}_2$: mediation effect attributable to the binary change of the mediator from zero to a non-zero status}\\
\multicolumn{9}{l}{NIE: total mediation effect}\\[1ex]
\end{tabular}
}
\end{table}

\subsection{Data with ZIP mediators}
In the ZIP data, the independent variable $X$ was generated from a standard normal distribution. $\text{NIE}_1$, $\text{NIE}_2$, and NIE were calculated for $X$ increasing from 0 to 1. The results (see Table \ref{table_simzip}) showed reasonably good performance for the proposed method on bias and CP of the 95\% CI. For the proposed method, AIC selected the ZIP model for over 93\% of the random samples while ZINB model was chosen for the rest. A Poisson-distributed $M$ and a binary $1_{(M>0)}$ were specified for the MSM method, and it had a similar pattern to that in Table \ref{table_simzinb} showing large biases and undercoverage of the 95\% CI. 

When zeros were only generated by a Poisson distribution (i.e., no excessive zeros or false zeros), the proposed method still performed better than MSM in terms of biases although MSM also had a reasonably good performance (see Table S2 in the supplementary materials).

The results for the empirical approach are presented in Tables S5 and S7 in the supplementary materials, and it showed a similar pattern to that of the ZINB mediators in the above section. When the mediator is zero-inflated, the results (see Table S5) showed large biases and very low CP in all scenarios with different percentages of zeros. When the mediator is not zero-inflated (i.e., no excessive or false zeros), the results (see Table S7) improved substantially. 

\begin{table}
\caption{Simulation results for ZIP data when $X$ changes from $x_1=0$ to $x_2=1$ from the final model chosen by AIC versus MSM. Mean SE is the mean of estimated standard errors. Bias is the mean estimates minus true values, and percent of bias is the bias as a percentage of true values. CP denotes the empirical coverage probability of the 95\% CI.} 
\label{table_simzip}
\centering
\resizebox{14cm}{!}{
\begin{tabular}{c c c c c c c c c} \\
\hline
Total zeros (\%) & Method & Mediation & True & Mean & Mean & Bias & Percent of Bias & CP\\
about half are false zeros& & Effect && Estimate & SE && (\%) & (\%)\\
\hline\hline\\
\multirow{1}{3em}{$\sim 30$} & Proposed & $\text{NIE}_1$ & -0.57 & -0.57 & 0.36 & 0.00 & 0.53 & 97.00 \\ 
&&$\text{NIE}_2$ & -0.28 & -0.27 & 0.15 & 0.01 & -2.90 & 93.00 \\ 
&&NIE & -0.84 & -0.84 & 0.41 & 0.01 & -0.71 & 93.00 \\ \\
&MSM&NIE  & -0.84 & 0.02 & 0.29 & 0.86 & -102.14 & 13.00 \\ 
\hline \\
\multirow{1}{3em}{$\sim 50$} & Proposed & $\text{NIE}_1$ & -0.45 & -0.48 & 0.33 & -0.03 & 7.33 & 97.00 \\ 
&&$\text{NIE}_2$ & -0.58 & -0.53 & 0.25 & 0.06 & -9.79 & 93.00 \\ 
&&NIE  & -1.03 & -1.01 & 0.46 & 0.02 & -2.33 & 96.00 \\   \\
&MSM&NIE  & -1.03 & 0.22 & 0.25 & 1.26 & -121.61 & 0.00 \\ 
\hline\\
\multirow{1}{3em}{$\sim 60$} & Proposed & $\text{NIE}_1$ & -0.38 & -0.37 & 0.32 & 0.01 & -3.14 & 93.00 \\ 
&&$\text{NIE}_2$  & -0.72 & -0.75 & 0.30 & -0.02 & 3.18 & 91.00 \\ 
&&NIE & -1.11 & -1.12 & 0.49 & -0.01 & 0.99 & 92.00 \\  \\
&MSM&NIE  & -1.11 & 0.28 & 0.23 & 1.39 & -125.68 & 0.00 \\ 
\hline\\
\multirow{1}{3em}{$\sim 70$} & Proposed & $\text{NIE}_1$ & -0.35 & -0.33 & 0.31 & 0.01 & -3.77 & 93.00 \\ 
&&$\text{NIE}_2$ & -0.78 & -0.83 & 0.34 & -0.04 & 5.73 & 93.00 \\ 
&&NIE & -1.13 & -1.16 & 0.52 & -0.03 & 2.83 & 92.00 \\  \\
&MSM&NIE  & -1.13 & 0.27 & 0.20 & 1.40 & -123.63 & 0.00 \\
\hline\\
\multirow{1}{3em}{$\sim 76$} & Proposed & $\text{NIE}_1$ & -0.31 & -0.30 & 0.30 & 0.01 & -3.50 & 96.00 \\ 
&&$\text{NIE}_2$ & -0.83 & -0.77 & 0.38 & 0.06 & -7.61 & 95.00 \\ 
&&NIE  & -1.14 & -1.07 & 0.55 & 0.07 & -6.48 & 99.00 \\ \\
&MSM&NIE & -1.14 & 0.22 & 0.18 & 1.36 & -119.00 & 0.00 \\ [1ex]
\hline
\multicolumn{9}{l}{Proposed: proposed method for ZILoN, ZINB, and ZIP mediators chosen by AIC}\\
\multicolumn{9}{l}{MSM: standard mediation analysis approach using marginal structural models}\\
\multicolumn{9}{l}{$\text{NIE}_1$: mediation effect attributable to the change in the mediator on the numerical scale}\\
\multicolumn{9}{l}{$\text{NIE}_2$: mediation effect attributable to the binary change of the mediator from zero to a non-zero status}\\
\multicolumn{9}{l}{NIE: total mediation effect}\\[1ex]
\end{tabular}
}
\end{table}

\section{Real data application - RANN Study}\label{sc:application}
The proposed mediation approach was applied to the motivating example described in section \ref{sc:moti} to examine the mediation effects of WMH on the potential causal pathway from aging to change in cognitive function. We analyzed four measures of cognitive function, RAs, as the dependent variable (i.e., $Y$): fluid reasoning, episodic memory, vocabulary, and perceptual speed composite score based on neuropsychological batteries. There are six potential mediators (i.e., $M$): regional WMH in five lobes of the brain including frontal, cingulate, occipital, temporal, parietal lobes, and total WMH. The proportion of zeros for the potential mediator variables ranges from 31\% to 76\%. The independent variable is age (i.e., $X$). The following model was used for all dependent variables:
\begin{align*}
&Y=\beta_0+\beta_1M+\beta_2 1_{(M>0)}+\beta_3X+\beta_4X1_{(M>0)}+\epsilon,
\end{align*}
which is slightly different than equation (\ref{ymodel}) since the interaction between $X$ and $M$ is not included here because they are both continuous variables. The proposed method was applied for each combination of outcome and mediator variables. The proposed models fits models with ZILoN, ZINB, and ZIP mediators on the real data respectively, and AIC was employed to decide the final mediation model. We compared our model with the MSM method. We used the R package ``CMAverse'' \cite{Shi2021} to implement the MSM method with a mediator $M$ and a binary mediator $1_{(M>0)}$. Once the final mediation model was determined, the chosen distribution was specified for the mediator $M$ in the MSM method. For example, if the final mediation model was chosen to be the model with ZINB mediators in the proposed method, a NB-distributed $M$ and a binary $1_{(M>0)}$ would be supplied to the MSM method. Significant results are shown in Table \ref{table_realdata2} where NIE$_1$, NIE$_2$, and NIE along with their 95\% CIs were estimated for age increasing from 50 to 70 years old.

We used 0.05 as the significance level and the results are presented before any multiple testing adjustment. Our method showed that the occipital WMH significantly mediates the effect of aging on memory through NIE$_1$ ($p=0.03$), but not the total mediation effect NIE ($p=0.06$) or NIE$_2$ ($p=0.61$). MSM did not show significance for this mediation effect. 
Both of our method and MSM method revealed that frontal WMH, parietal WMH, and total WMH have significant total mediation effects on the perceptual speed processing ability. Furthermore, our model showed that NIE$_2$ is not significant for those three mediators and NIE$_1$ is significant for the parietal and total WMH. 
The MSM method showed that cingulate WMH has a significant total mediation effect on the perceptual speed processing ability while our method did not show significance for cingulate WMH. Our model generated more significant findings than the previous report.\cite{moura2019relationship} 

\begin{table}
\caption{Real data analysis results for the estimation of mediation effects when age increases from 50 to 70 years old from the final model chosen by AIC versus MSM.} 
\label{table_realdata2}
\centering
\resizebox{14cm}{!}{
\begin{tabular}{c c c c c c c c} \\
\hline
Outcome & Mediator & Method & Mediation & Estimate & SE & 95\% CI & p-value\\
&&& Effect &&&&\\
\hline\hline\\
Memory & Occipital &MedZINB & $\text{NIE}_1$ & -0.07 & 0.03 & (-0.13, -0.01) & $0.03^*$ \\ 
&&&$\text{NIE}_2$  & -0.02 & 0.05 & (-0.11, 0.07) & 0.61 \\ 
&&&NIE  & -0.09 & 0.05 & (-0.19, 0) & 0.06 \\ \\
&&MSM&NIE  & -0.08 & 0.05 & (-0.19, 0) & 0.08 \\ 
\hline\\
Perceptual speed & Frontal &MedZILoN & $\text{NIE}_1$ &-0.03 &0.02 &(-0.06, 0.00) &0.07\\
&&&$\text{NIE}_2$ &-0.06 &0.04 &(-0.14, 0.02) &0.13\\
&&&NIE &-0.09 &0.04 &(-0.17, -0.01) &$0.03^*$\\\\
&&MSM$^+$&NIE & -0.10 & 0.06 & (-0.25, -0.03) & $0.01^*$\ \\ 
\hline\\
Perceptual speed & Cingulate &MedZINB & $\text{NIE}_1$  & -0.02 & 0.01 & (-0.04, 0.01) & 0.23 \\ 
&&&$\text{NIE}_2$  & -0.05 & 0.04 & (-0.13, 0.03) & 0.23 \\ 
&&&NIE & -0.07 & 0.04 & (-0.14, 0.01) & 0.09 \\ \\
&&MSM&NIE  & -0.06 & 0.03 & (-0.13, 0) & $0.05^*$ \\ 
\hline\\
Perceptual speed & Parietal &MedZINB & $\text{NIE}_1$  & -0.06 & 0.03 & (-0.11, 0) & $0.04^*$ \\ 
&&&$\text{NIE}_2$  & -0.09 & 0.06 & (-0.2, 0.02) & 0.12 \\ 
&&&NIE  & -0.15 & 0.06 & (-0.26, -0.04) & $0.01^*$ \\  \\
&&MSM&NIE  & -0.13 & 0.05 & (-0.26, -0.05) & $0.00^*$ \\ 
\hline\\
Perceptual speed & Total &MedZILoN & $\text{NIE}_1$ & -0.06 & 0.03 & (-0.11, -0.01) & $0.03^*$ \\ 
&&&$\text{NIE}_2$ & -0.04 & 0.04 & (-0.13, 0.04) & 0.31 \\
&&&NIE  & -0.10 & 0.05 & (-0.19, -0.01) & $0.04^*$ \\ \\
&&MSM$^+$&NIE & -0.09 & 0.05 & (-0.21, -0.02) & $0.01^*$ \\[1ex]
\hline
\multicolumn{8}{l}{MedZINB: proposed mediation model for zero-inflated negative binomial mediators}\\
\multicolumn{8}{l}{MedZILoN: proposed mediation model for zero-inflated log-normal mediators}\\
\multicolumn{8}{l}{MSM: standard mediation analysis approach using marginal structural models}\\
\multicolumn{8}{l}{$\text{NIE}_1$: mediation effect attributable to the change in the mediator on the numerical scale}\\
\multicolumn{8}{l}{$\text{NIE}_2$: mediation effect attributable to the binary change of the mediator from zero to a non-zero status}\\
\multicolumn{8}{l}{NIE: total mediation effect}\\
\multicolumn{8}{l}{$^+$MSM method with a NB-distributed $M$ and a binary $1_{(M>0)}$ as the existing R package is not applicable}\\[1ex]
\end{tabular}
}
\end{table}

\section{Discussion}\label{sc:diss}
We proposed a novel approach to estimate and test the mediation effects of zero-inflated mediators that are commonly seen in biomedical studies, especially in neuroscience and brain imaging studies. Mixed-type two-part distributions were considered for the mediator to account for its zero-inflated structure. The total mediation effect is decomposed into two components.
The first component is the mediation effect attributable to the numeric change of the mediator summing two causal pathways bypassing $M$: $X\rightarrow M\rightarrow Y$ and $X\rightarrow M\rightarrow 1_{(M>0)}\rightarrow Y$, denoted NIE$_1$. The second component is the mediation effect attributable to the binary change between zero and non-zero status of the mediator on the causal pathway bypassing only $1_{(M>0)}$: $X\rightarrow 1_{(M>0)}\rightarrow Y$, denoted NIE$_2$. The problem of false zeros is addressed by employing a probability mechanism for observing such zeros. Compared with standard causal mediation analysis methods like MSM method, our model has better performance as demonstrated by an extensive simulation study. Potential confounders or covariates can be easily adjusted in the regression equations of the proposed approach. And we have made an R package called ``MAZE'' for implementing the method and it can be installed through a github website (``https://github.com/meilinjiang/MAZE'').

The mechanism of observing false zeros in equation (\ref{zeroMecha}) is an important part of the model to account for false zeros. Our approach has the flexibility to accommodate different mechanisms. For example, one could specify the mechanism using $\exp{(-\eta^2\sqrt{M})}$ instead of $\exp{(-\eta^2M)}$ if needed. In the case of multiple mechanisms being used, model selection approaches such as AIC and BIC can be employed to choose a mechanism among multiple candidates. Due to the zero-inflated structure, it is challenging and computationally demanding to deal with joint distributions of zero-inflated mediators. A possible extension for the future development of our approach is to develop methodology that allows for multiple or high-dimensional mediators to be included in the mediation model.  

We focused on continuous outcomes for $Y$ in this paper, but it can be easily extended to general outcomes. The mediation effect can be decomposed in the same way as it is in equation (\ref{decomposeNIE}) for a generalized linear model for $Y$ in equation (\ref{ymodel}). However, the formulas for NIE, NIE$_1$ and NIE$_2$ may not always have explicit expressions like those for a linear regression model in equation (\ref{ymodel}). For example, when $Y$ is binary and a logistic regression model is used in equation (\ref{ymodel}), the mediation effect is usually defined in terms of odds ratio and no longer has a general closed-form expression.\cite{VanderWeele2010} We will proceed with this line of development in future research.

\section*{Acknowledgments}
This research work was partly supported by NIH grants: R01GM123014, P01AA029543, R01AG026158, R01AG038465, and R01AG062578.

\subsection*{Data availability}
Research data are not shared.

\subsection*{Conflict of interest}
The authors declare no potential conflict of interests.

\section*{Supporting information}
The supplementary materials can be found online in the Supporting Information section at the end of this article.

\appendix
\section{Detailed derivation for estimation with EM algorithm} \label{sc:appendix_a}
\subsection{Log-likelihood function}
Using the observed data variable vector $(y_i,m_i^*,r_i,c_i,x_i)$ for the $i$th subject, the log-likelihood function can be derived as follows:
\begin{align*}
\ell&=\log\Big(\prod_{i=1}^Nf(y_i,m_i^*,r_i,c_i|x_i)\Big)\\
&=\log{\Bigg\{\prod_{i=1}^N\Big[f(y_i,m_i^*,r_i|x_i,c_i=0)P(C_i=0)\Big]^{1_{(c_i=0)}}\Big[f(y_i,m_i^*,r_i|x_i,c_i=1)P(C_i=1)\Big]^{1_{(c_i=1)}}\Bigg\}}\\
&=\log{\Bigg\{\prod_{i=1}^N\Big[\Delta_if(y_i,m_i^*,r_i|x_i,c_i=0)\Big]^{1_{(c_i=0)}}\Big[(1-\Delta_i)f(y_i,m_i^*,r_i|x_i,c_i=1)\Big]^{1_{(c_i=1)}}\Bigg\}}\\
&=\sum_{i=1}^N\Bigg\{1_{(c_i=0)}\Big[\log(\Delta_i)+\log(f(y_i,m_i^*,r_i|x_i,c_i=0))\Big]+1_{(c_i=1)}\Big[\log(1-\Delta_i)+\log(f(y_i,m_i^*,r_i|x_i,c_i=1))\Big]\Bigg\},
\end{align*}
Let $\ell_{i0}=\log{\big(f(y_i,m_i^*,r_i|x_i,c_i=0)\big)}$ and $\ell_{i1}=\log{\big(f(y_i,m_i^*,r_i|x_i,c_i=1)\big)}$. Then
\begin{align*}
&\ell=\sum_{i=1}^N\Bigg\{1_{(c_i=0)}\Big[\log(\Delta_i)+\ell_{i0}\Big]+1_{(c_i=1)}\Big[\log(1-\Delta_i)+\ell_{i1}\Big]\Bigg\}.
\end{align*}
After dividing all subjects into two groups by whether $m_i^*$ is zero, the complete log-likelihood function can be written as:
\begin{align*}
\ell&=\sum_{i\in\text{group1}}1_{(c_i=1)}\Big[\log(1-\Delta_i)+\ell_{i1}^1\Big]+\sum_{i\in\text{group2}}\Bigg\{1_{(c_i=0)}\Big[\log(\Delta_i)+\ell_{i0}^2\Big]+1_{(c_i=1)}\Big[\log(1-\Delta_i)+\ell_{i1}^2\Big]\Bigg\}\nonumber\\
&=\sum_{i\in\text{group1}}\Big[\log(1-\Delta_i)+\ell_{i1}^1\Big]+\sum_{i\in\text{group2}}\Bigg\{1_{(c_i=0)}\Big[\log(\Delta_i)+\ell_{i0}^2\Big]+1_{(c_i=1)}\Big[\log(1-\Delta_i)+\ell_{i1}^2\Big]\Bigg\},
\end{align*}
where $\ell_{i1}^1$ denotes $\ell_{i1}$ in group 1, and $\ell_{i0}^2$ and $\ell_{i1}^2$ denote $\ell_{i0}$ and $\ell_{i1}$ in group 2 respectively. 

\subsection{Expectation step (E step)}
Let $\tau_{ik}(\Theta^0)=E_{c_i|y_i,m_i^*,r_i,x_i,\Theta^0}\big(1_{(c_i=k)}\big)$ for $k=0,1$ in the conditional expectation $Q(\Theta|\Theta^0)$. By using the Bayes formula, we have
\begin{align*}
\tau_{i0}(\Theta^0)&=E\big(1_{(c_i=0)}|y_i,m_i^*=0,r_i=0,x_i,\Theta^0\big)\\
&=P\big(C_i=0|y_i,m_i^*=0,r_i=0,x_i,\Theta^0\big)\\
&=\frac{f(y_i,m_i^*=0,r_i=0,c_i=0|x_i,\Theta^0)}{f(y_i,m_i^*=0,r_i=0|x_i,\Theta^0)}\\
&=\frac{f(y_i,m_i^*=0,r_i=0|c_i=0,x_i,\Theta^0)P(C_i=0|x_i,\Theta^0)}{\sum_{k=0}^1f(y_i,m_i^*=0,r_i=0|c_i=k,x_i,\Theta^0)P(C_i=k|x_i,\Theta^0)}\\
&=\frac{\Delta_i\exp{(\ell_{i0}^2)}}{\Delta_i\exp{(\ell_{i0}^2)}+(1-\Delta_i)\exp{(\ell_{i1}^2)}}\Bigg|_{\hspace{0.1cm}\text{evaluated at}\hspace{0.1cm} \Theta^0},\\
\tau_{i1}(\Theta^0)&=E\big(1_{(c_i=1)}|y_i,m_i^*=0,r_i=0,x_i,\Theta^0\big)\\
&=1-\tau_{i0}(\Theta^0).
\end{align*}
Finally we get 
\begin{align*}
Q(\Theta|\Theta^0)&=E(\ell|y_i,m_i^*,r_i,x_i,\Theta^0)\nonumber\\
&=\sum_{i\in\text{group1}}\Big[\log(1-\Delta_i)+\ell_{i1}^1\Big]+\sum_{i\in\text{group2}}\Bigg\{\tau_{i0}(\Theta^0)\Big[\log(\Delta_i)+\ell_{i0}^2\Big]+\tau_{i1}(\Theta^0)\Big[\log(1-\Delta_i)+\ell_{i1}^2\Big]\Bigg\}.
\end{align*}

\section{Detailed derivation for the log-likelihood function of models with ZILoN mediators} \label{sc:appendix_b}
For group 1 consisting of subjects with positive mediator values (i.e., $r_i=1$), the log-likelihood contribution from the $i$th individual can be calculated as:
\begin{align*}
\ell_{i1}^1&=\log{\big(f(y_i,m_i^*,r_i=1|x_i,c_i=1)\big)}\\
&=\log(f(y_i, r_i=1|m_i^*,x_i,c_i=1)f_M(m_i^*|x_i,c_i=1))\\
&=\log(f(y_i|m_i^*,x_i,c_i=1)P(R_i=1|m_i^*,x_i,c_i=1)f_M(m_i^*|x_i,c_i=1))\\
&=\log(f(y_i|m_i^*,x_i))+\log(P(R_i=1|m_i^*))+\log(f_M(m_i^*|x_i,c_i=1))\\
&=\log(f(y_i|m_i^*,x_i))+\log(P(M_i^*>0|m_i^*))+\log(f_M(m_i^*|x_i,c_i=1))\\
&=-\log(\delta)-\frac{(y_i-\beta_0-\beta_1m_i^*-\beta_2-(\beta_3+\beta_4)x_i-\beta_5x_im_i^*)^2}{2\delta^2}+\log\Big[1-1_{(m_i^*\le B)}\exp{(-\eta^2m_i^*)}\Big]\\
&\hspace{0.4cm}-\log(m_i^*\sigma)-\frac{(\log (m_i^*)-\mu_i)^2}{2\sigma^2}-\log{(2\pi)}.
\end{align*}
The subjects in group 2 have zeroed observed mediator values (i.e., $r_i=0$). The log-likelihood contribution of those individuals in group 2 with a true zero ($m_i^*=m_i=0$) is
\begin{align*}
\ell_{i0}^2 &= \log(f(y_i,m_i^*=0,r_i=0|x_i,c_i=0))\\
&= \log(f(y_i|m_i^*=0,x_i,c_i=0)P(M_i^*=0,R_i=0|x_i,c_i=0))\\
&= \log(f(y_i|m_i^*=0,x_i)P(R_i=0|m_i^*=0,x_i,c_i=0)P(M_i^*=0|x_i,c_i=0))\\
&= \log(f(y_i|m_i^*=0,x_i)P(R_i=0|m_i^*=0)P(M_i^*=0|x_i,c_i=0))\\
&= \log(f(y_i|m_i^*=0,x_i))\\
&=-\log(\delta)-\frac{(y_i-\beta_0-\beta_3x_i)^2}{2\delta^2}-0.5\log{(2\pi)}.
\end{align*}
The remainder of the group 2 subjects are individuals with a false zero ($m_i>0$ and $m_i^*=0$) with a log-likelihood contribution:
\begin{align*}
\ell_{i1}^2&=\log(f(y_i,m_i^*=0,r_i=0|x_i,c_i=1))\\
&=\log\bigg(\int\limits_0^B f(y_i, r_i=0|m,x_i,c_i=1)dF_M(m|x_i,c_i=1)\bigg)\\
&=\log\bigg(\int\limits_0^B f(y_i|m,x_i,c_i=1) P(R_i=0|m,x_i,c_i=1)f_M(m|x_i,c_i=1)dm\bigg)\\
&=\log\bigg(\int\limits_0^B f(y_i|m,x_i) P(R_i=0|m)\phi(m;\mu_i,\sigma)dm\bigg)\\
&=\log\Bigg\{\int\limits_0^B \frac{1}{\delta\sqrt{2\pi}}\exp{\bigg[-\frac{(y_i-\beta_0-\beta_1m-\beta_2-(\beta_3+\beta_4)x_i-\beta_5x_im)^2}{2\delta^2}\bigg]}\\
&\hspace{1.8cm}\cdot \exp{(-\eta^2m)}\frac{1}{m\sigma\sqrt{2\pi}}\exp\bigg[-\frac{(\log (m)-\mu_i)^2}{2\sigma^2}\bigg]dm\Bigg\}\\
&=\log\Bigg(\frac{1}{2\pi\delta\sigma}\int\limits_0^B h_i(m)dm\Bigg),
\end{align*}
where 
\begin{align*}
h_i(m)=\frac{1}{m}\exp\bigg(-\frac{(\log (m)-\mu_i)^2}{2\sigma^2}-\frac{(y_i-\beta_0-\beta_1m-\beta_2-(\beta_3+\beta_4)x_i-\beta_5x_im)^2}{2\delta^2}-\eta^2m\bigg).
\end{align*}

\bibliography{../draft_ref_220920}

\begin{thebibliography}{10}
\providecommand \doibase [0]{http://dx.doi.org/}%

\bibitem{mackinnon2007mediation}
MacKinnon DP, Fairchild AJ, Fritz MS. Mediation analysis. {\it Annu. Rev.
  Psychol.} 2007\string; 58\string: 593--614.

\bibitem{VanderWeele2009}
VanderWeele TJ. Marginal Structural Models for the Estimation of Direct and
  Indirect Effects. {\it Epidemiology} 2009\string; 20(1)\string: 18--26.
\newblock \href {\doibase 10.1097/ede.0b013e31818f69ce} {doi:
  10.1097/ede.0b013e31818f69ce}

\bibitem{Imai2010}
Imai K, Keele L, Tingley D. A general approach to causal mediation analysis..
  {\it Psychological Methods} 2010\string; 15(4)\string: 309--334.
\newblock \href {\doibase 10.1037/a0020761} {doi: 10.1037/a0020761}

\bibitem{Judd1981}
Judd CM, Kenny DA. Process Analysis: Estimating Mediation in Treatment
  Evaluations. {\it Evaluation Review} 1981\string; 5(5)\string: 602--619.
\newblock \href {\doibase 10.1177/0193841x8100500502} {doi:
  10.1177/0193841x8100500502}

\bibitem{Sobel1982}
Sobel ME. Asymptotic Confidence Intervals for Indirect Effects in Structural
  Equation Models. {\it Sociological Methodology} 1982\string; 13\string: 290.
\newblock \href {\doibase 10.2307/270723} {doi: 10.2307/270723}

\bibitem{Baron1986}
Baron RM, Kenny DA. The moderator{\textendash}mediator variable distinction in
  social psychological research: Conceptual, strategic, and statistical
  considerations.. {\it Journal of Personality and Social Psychology}
  1986\string; 51(6)\string: 1173--1182.
\newblock \href {\doibase 10.1037/0022-3514.51.6.1173} {doi:
  10.1037/0022-3514.51.6.1173}

\bibitem{Robins1992}
Robins JM, Greenland S. Identifiability and Exchangeability for Direct and
  Indirect Effects. {\it Epidemiology} 1992\string; 3(2)\string: 143--155.
\newblock \href {\doibase 10.1097/00001648-199203000-00013} {doi:
  10.1097/00001648-199203000-00013}

\bibitem{Pearl2001}
Pearl J. Direct and Indirect Effects. In: UCLA. ; 2001; San Francisco: Morgan
  Kaufmann\string: 411–420.

\bibitem{Vanderweele2015}
Vanderweele T. {\it Explanation in Causal Inference: Methods for Mediation and
  Interaction}.
\newblock New York: Oxford University Press .
\newblock 2015.

\bibitem{Zheng2014}
Zheng C, Zhou XH. Causal mediation analysis in the multilevel intervention and
  multicomponent mediator case. {\it Journal of the Royal Statistical Society:
  Series B (Statistical Methodology)} 2014\string; 77(3)\string: 581--615.
\newblock \href {\doibase 10.1111/rssb.12082} {doi: 10.1111/rssb.12082}

\bibitem{Guo2018}
Guo Z, Small DS, Gansky SA, Cheng J. Mediation analysis for count and
  zero-inflated count data without sequential ignorability and its application
  in dental studies. {\it Journal of the Royal Statistical Society: Series C
  (Applied Statistics)} 2018\string; 67(2)\string: 371--394.
\newblock \href {\doibase 10.1111/rssc.12233} {doi: 10.1111/rssc.12233}

\bibitem{VanderWeele2016}
VanderWeele TJ. Mediation Analysis: A Practitioner{\textquotesingle}s Guide.
  {\it Annual Review of Public Health} 2016\string; 37(1)\string: 17--32.
\newblock \href {\doibase 10.1146/annurev-publhealth-032315-021402} {doi:
  10.1146/annurev-publhealth-032315-021402}

\bibitem{wardlaw2015white}
Wardlaw JM, Vald{\'e}s~Hern{\'a}ndez MC, Mu{\~n}oz-Maniega S. What are white
  matter hyperintensities made of? Relevance to vascular cognitive impairment.
  {\it Journal of the American Heart Association} 2015\string; 4(6)\string:
  e001140.

\bibitem{habes2016white}
Habes M, Erus G, Toledo JB, et al. White matter hyperintensities and imaging
  patterns of brain ageing in the general population. {\it Brain} 2016\string;
  139(4)\string: 1164--1179.

\bibitem{merino2019white}
Merino JG. White matter hyperintensities on magnetic resonance imaging: what is
  a clinician to do?. In: . 94. Elsevier. ; 2019\string: 380--382.

\bibitem{lee2016white}
Lee S, Viqar F, Zimmerman ME, et al. White matter hyperintensities are a core
  feature of Alzheimer's disease: evidence from the dominantly inherited
  Alzheimer network. {\it Annals of neurology} 2016\string; 79(6)\string:
  929--939.

\bibitem{lee2018white}
Lee S, Zimmerman ME, Narkhede A, et al. White matter hyperintensities and the
  mediating role of cerebral amyloid angiopathy in dominantly-inherited
  Alzheimer’s disease. {\it PLoS One} 2018\string; 13(5)\string: e0195838.

\bibitem{moura2019relationship}
Moura AR, Lee S, Habeck C, Razlighi Q, Stern Y. The relationship between white
  matter hyperintensities and cognitive reference abilities across the life
  span. {\it Neurobiology of aging} 2019\string; 83\string: 31--41.

\bibitem{Francois2012}
Francois M, Peter C, Gordon F. Dealing with excess of zeros in the statistical
  analysis of magnetic resonance imaging lesion count in multiple sclerosis.
  {\it Pharmaceutical Statistics} 2012\string; 11(5)\string: 417--424.
\newblock \href {\doibase 10.1002/pst.1529} {doi: 10.1002/pst.1529}

\bibitem{wuMediation}
Wu Q, O'Malley J, Liyanage J, et al. MedZIM: Mediation analysis for
  Zero-Inflated Mediators with applications to microbiome data. {\it
  arXiv:1906.09175v3 [stat.ME]} 2021.

\bibitem{Stern2009}
Stern Y. Cognitive reserve. {\it Neuropsychologia} 2009\string; 47(10)\string:
  2015--2028.
\newblock \href {\doibase 10.1016/j.neuropsychologia.2009.03.004} {doi:
  10.1016/j.neuropsychologia.2009.03.004}

\bibitem{Stern2014}
Stern Y, Habeck C, Steffener J, et al. The Reference Ability Neural Network
  Study: Motivation, design, and initial feasibility analyses. {\it
  {NeuroImage}} 2014\string; 103\string: 139--151.
\newblock \href {\doibase 10.1016/j.neuroimage.2014.09.029} {doi:
  10.1016/j.neuroimage.2014.09.029}

\bibitem{Stern2018}
Stern Y, Gazes Y, Razlighi Q, Steffener J, Habeck C. A task-invariant cognitive
  reserve network. {\it {NeuroImage}} 2018\string; 178\string: 36--45.
\newblock \href {\doibase 10.1016/j.neuroimage.2018.05.033} {doi:
  10.1016/j.neuroimage.2018.05.033}

\bibitem{Salthouse2009}
Salthouse TA. Decomposing age correlations on neuropsychological and cognitive
  variables. {\it Journal of the International Neuropsychological Society}
  2009\string; 15(5)\string: 650--661.
\newblock \href {\doibase 10.1017/s1355617709990385} {doi:
  10.1017/s1355617709990385}

\bibitem{Salthouse2015}
Salthouse TA, Habeck C, Razlighi Q, Barulli D, Gazes Y, Stern Y. Breadth and
  age-dependency of relations between cortical thickness and cognition. {\it
  Neurobiology of Aging} 2015\string; 36(11)\string: 3020--3028.
\newblock \href {\doibase 10.1016/j.neurobiolaging.2015.08.011} {doi:
  10.1016/j.neurobiolaging.2015.08.011}

\bibitem{Prins2015}
Prins ND, Scheltens P. White matter hyperintensities, cognitive impairment and
  dementia: an update. {\it Nature Reviews Neurology} 2015\string;
  11(3)\string: 157--165.
\newblock \href {\doibase 10.1038/nrneurol.2015.10} {doi:
  10.1038/nrneurol.2015.10}

\bibitem{Steen2017}
Steen J, Loeys T, Moerkerke B, Vansteelandt S. Flexible Mediation Analysis With
  Multiple Mediators.. {\it American journal of epidemiology} 2017\string;
  186\string: 184--193.
\newblock \href {\doibase 10.1093/aje/kwx051} {doi: 10.1093/aje/kwx051}

\bibitem{TerHorst1986}
TerHorst H. On Stieltjes integration in Euclidean space. {\it Journal of
  Mathematical Analysis and Applications} 1986\string; 114(1)\string: 57--74.
\newblock \href {\doibase 10.1016/0022-247x(86)90066-1} {doi:
  10.1016/0022-247x(86)90066-1}

\bibitem{Efron1986}
Efron B, Tibshirani R. Bootstrap Methods for Standard Errors, Confidence
  Intervals, and Other Measures of Statistical Accuracy. {\it Statistical
  Science} 1986\string; 1(1).
\newblock \href {\doibase 10.1214/ss/1177013815} {doi: 10.1214/ss/1177013815}

\bibitem{multipleMedi14}
VanderWeele T, Vansteelandt S. Mediation Analysis with Multiple Mediators. {\it
  Epidemiologic Methods} 2014\string; 2(1)\string: 95-115.

\bibitem{Oakes1999}
Oakes D. Direct calculation of the information matrix via the EM. {\it Journal
  of the Royal Statistical Society: Series B (Statistical Methodology)}
  1999\string; 61(2)\string: 479--482.

\bibitem{Shi2021}
Shi B, Choirat C, Coull BA, VanderWeele TJ, Valeri L. {CMAverse}: A Suite of
  Functions for Reproducible Causal Mediation Analyses. {\it Epidemiology}
  2021\string; 32(5)\string: e20--e22.
\newblock Available from https://bs1125.github.io/CMAverse/\href {\doibase
  10.1097/ede.0000000000001378} {doi: 10.1097/ede.0000000000001378}

\bibitem{VanderWeele2010}
VanderWeele TJ, Vansteelandt S. Odds Ratios for Mediation Analysis for a
  Dichotomous Outcome. {\it American Journal of Epidemiology} 2010\string;
  172(12)\string: 1339--1348.
\newblock \href {\doibase 10.1093/aje/kwq332} {doi: 10.1093/aje/kwq332}

\end{thebibliography}

\end{document}